\documentclass[twocolumn,pra,showpacs,superscriptaddress]{revtex4}

\usepackage{amsfonts}
\usepackage{amsmath}
\usepackage{amssymb}
\usepackage[dvips]{graphicx}

\begin{document}

\author{L. H. Pedersen}
\email{lhp@phys.au.dk} \affiliation{Lundbeck Foundation Theoretical
Center for Quantum System Research, Department of Physics and
Astronomy, University of Aarhus, DK-8000 \AA rhus C, Denmark}
\author{C. Rangan}
\email{rangan@uwindsor.ca} \affiliation{Department of Physics,
University of Windsor, Ontario N9B 3P4, Canada}

\title{Controllability and universal three-qubit quantum computation with trapped electron states}

\begin{abstract}
We show how to control and perform universal three-qubit quantum
computation with trapped electron quantum states.  The three qubits
are the electron spin, and the first two quantum states of the
cyclotron and axial harmonic oscillators. We explicitly show how
universal 3-qubit gates can be performed.  As an example of a quantum
algorithm, we outline the implementation of the 3-qubit Deutsch-Jozsa
algorithm in this system.
\end{abstract}
 \maketitle

\section{Introduction}
\noindent Since the eighties, much effort has been put into the
study of quantum computers, and various proposals for the physical
implementation have been put forward in very different fields such
as those of cold trapped ions \cite{CZ95}, quantum dots
\cite{PBCDZRZ03}, NMR \cite{GC97} and neutral atoms \cite{SW05}.
Experimental implementations have, however, been few and limited to
a low number of qubits. It is therefore interesting to study
alternative candidates for quantum computers.\\
\indent In the present work we study a scheme put forward for
instance in \cite{MMT99} in which an electron trapped in a Penning
trap is considered a possible realization of quantum logic.
Advantages such as an almost complete absence of decoherence
mechanisms and already obtained good experimental accuracy suggests
this to be an interesting direction of study.  We show that it is
possible to perform universal three-qubit quantum computation in
this system, and outline the implementation of the three-qubit
Deutsch-Jozsa
algorithm in this system.\\
\indent In our scheme, quantum information is stored in an internal
degree of freedom (spin) and two external degrees of freedom
(cyclotron and axial motion) of the trapped electron. For two
qubits, proposals have already been put forward in
\cite{MMT99} and \cite{SM05}. Here, instead of restricting ourselves
to two qubits, we exploit an additional degree of freedom in a single electron and consider three
qubits.\\
\indent Note that although universal quantum computing for three
qubits of a single electron has already been demonstrated
theoretically in \cite{CMT02}, this proposal relied on small
relativistic effects which lead to anharmonicities for the
cyclotron oscillator.  In our scheme, we treat the cyclotron
oscillator as essentially harmonic (to experimental resolution)
and we use a traveling field that is experimentally
well-established \cite{WABBDHHQSVVV06}.  This scheme is distinctly different from schemes in
alternative trap configurations such as in Refs. \cite{CMT03} and
\cite{CMT04}, where a linear array of Penning traps was
investigated, and in \cite{CGMT05}, where an array of planar Penning
traps was considered.\\
\indent In the following section, we briefly describe the physical
system and the theoretical model that describes it.  Then we
describe the control mechanisms and show the control equations.
Section 4 demonstrates that universal 3-qubit quantum computation is
possible using these controls.  Subsequently, we present an explicit
example of the 3-qubit Deutsch-Jozsa algorithm that can be
implemented.  In the last section, we discuss the experimental
challenges to the implementation of this scheme.

\section{The system}
\noindent The physical system consisting of a single electron
trapped in a Penning trap has already been described in some detail
in for instance \cite{CMT01}. To summarize, the electron experiences
a magnetic field $\mathbf{B}=B \hat{\mathbf{z}}$ and a static
quadrupole potential
\begin{equation*}
V = V_0 \frac{z^2-(x^2+y^2)/2}{2d^2}
\end{equation*}
where $V_0$ is the potential between the trap electrodes and $d$ is
a characteristic length of the trap \cite{BG86}. The Hamiltonian,
given by
\begin{equation*}
H_0 = \frac{1}{2m}\left(
\mathbf{p}-e\mathbf{A}_0\right)^2+eV-\boldsymbol{\mu}\cdot
\mathbf{B}
\end{equation*}
where $\mathbf{A}_0 = \frac{1}{2}\mathbf{B}\times \mathbf{r}$ and
$\boldsymbol{\mu} = \frac{ge\hbar}{4m}\boldsymbol{\sigma}$, can be
recast in terms of three harmonic oscillators \cite{BG86}
\begin{equation*}
H_0 = \hbar \omega'_c a^\dagger_c a_c + \hbar \omega_z a^\dagger_z
a_z - \hbar \omega_m a^\dagger_m a_m + \frac{\hbar}{2} \omega_s
\sigma_z
\end{equation*}
namely the cyclotron, axial and magnetron motions. The frequencies
are given by $\omega_c'=(\omega_c+\tilde\omega_c)/2$, $\omega_m =
(\omega_c-\tilde\omega_c)/2$, $\omega_z = \sqrt{\frac{eV_0}{md^2}}$,
$\tilde\omega_c = \sqrt{\omega_c^2-2\omega_z^2}$, $\omega_c =
\frac{|e|B}{m}$ and $\omega_s = \frac{g|e|B}{2m}$. In \cite{MMT99}
the spin and axial degrees of freedom are used as qubits whereas in
\cite{SM05} spin and the cyclotron oscillator constitute the qubits.
Here we use the spin of the electron and the axial and the cyclotron
oscillator states for the three qubits. The logical states
$|0\rangle$ and $|1\rangle$ corresponds to
$|\!\!\downarrow\,\rangle$ and $|\!\!\uparrow\,\rangle$ for the spin
and the Fock states $|0\rangle$ and $|1\rangle$ for the cyclotron
and the axial oscillators. The states are denoted by $|j n l\rangle$
where $j,n,l$ are the logical states for the spin, cyclotron and
axial states, respectively. The frequencies corresponding to the
qubit-energy splittings are of the order \textrm{GHz}, \textrm{GHz}
and \textrm{MHz}, respectively. See Fig. 1 for an energy-level
diagram.
\begin{figure}[b]
\includegraphics[width=10cm]{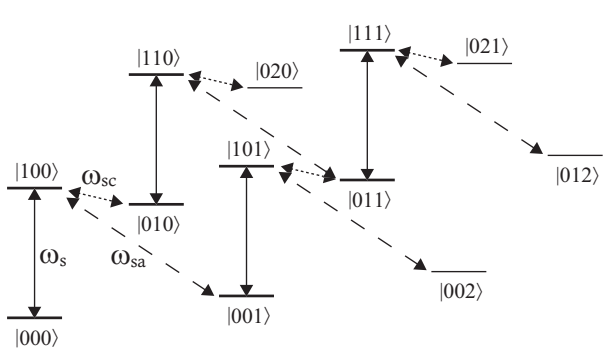}
\caption{{\small Energy-level diagram. States are denoted
$|jnl\rangle$, where $j$ is the spin state, $n$ the cyclotron state
and $l$ the axial state. Levels marked with bold are included in the
computational subspace. Solid, dashed and dotted lines indicate
spin, spin-axial and spin-cyclotron transitions, respectively.}}
\end{figure}

\section{Controlling the system}\label{control}

\noindent The spin qubit is the only qubit which can be controlled
directly. This is done using a small transverse magnetic field
\cite{CMT01}
\begin{equation*}
\mathbf{b}(t) = b\left( \cos(\omega t+\phi) \hat{x}+\sin(\omega
t+\phi) \hat{y}\right)
\end{equation*}
If $\omega$ is close to $\omega_s$ this leads to the following
Hamiltonian in the interaction picture
\begin{equation*}
H'_s = \frac{\hbar \Omega}{2}\left( \sigma_+ \exp(-i\phi)+\sigma_-
\exp(i\phi)\right)
\end{equation*}
where $\Omega = \frac{g|e|b}{2m}$. In the basis $|\!\downarrow\! n
l\rangle$, $|\!\uparrow\! n l\rangle$ this interaction implies the
evolution matrix
\begin{equation*}
U'_s = \begin{pmatrix} \cos(\frac{\Omega t}{2}) &
-ie^{i\phi}\sin(\frac{\Omega t}{2}) \\
-ie^{-i\phi}\sin(\frac{\Omega t}{2}) & \cos(\frac{\Omega t}{2})
\end{pmatrix}
\end{equation*}
subsequently referred to as a $p_s(\theta,\phi)$ pulse, where
$\theta = \Omega t$.\\
\indent The cyclotron and axial qubits
cannot be directly controlled. Cyclotron transitions can,
admittedly, be addressed by setting up a proper vector potential
\cite{CMT01}, but since the levels are equally spaced, population
will inevitably leave the computational
subspace~\cite{RanganPRL2004}.  To control the spin-axial transition
a traveling magnetic field is set up
\begin{equation*}
\mathbf{b}(t) = b\left( \cos(\omega t+\phi-kz) \hat{x}+\sin(\omega
t+\phi-kz) \hat{y}\right)
\end{equation*}
which leads to the following Hamiltonian in the interaction picture
\begin{equation*}
\begin{split}
H'_{sa} = &\frac{\hbar \Omega}{2} \bigl( \sigma_+
\exp(i(\eta(a^\dagger_z e^{i\omega_z t}+a_z e^{-i\omega_z t})-\Delta
\cdot t-\phi)) \\ &+\sigma_- \exp(-i(\eta(a^\dagger_z e^{i\omega_z
t}+a_z e^{-i\omega_z t})-\Delta \cdot t-\phi))\bigr)
\end{split}
\end{equation*}
where $\Omega = \frac{g|e|b}{2m}$, $\Delta = \omega-\omega_s$ and
$\eta = k\sqrt{\frac{\hbar}{2m\omega_z}}$. For $\omega = \omega_{sa}
= \omega_s-\omega_z$ only the levels $|\!\downarrow\! n
(l+1)\rangle$ and $|\!\uparrow\! n l\rangle$ are connected. In the
basis $|\!\downarrow\! n (l+1)\rangle$, $|\!\uparrow\! n l\rangle$
the evolution matrix is given by
\begin{equation}\label{Usa}
U'_{sa} = \begin{pmatrix} \cos(\frac{\theta_l}{2}) & -e^{i\phi}\sin(\frac{\theta_l}{2}) \\
e^{-i\phi}\sin(\frac{\theta_l}{2}) & \cos(\frac{\theta_l}{2})
\end{pmatrix}
\end{equation}
where $\theta_l = \Omega t \cdot \eta
e^{-\eta^2/2}\sqrt{\tfrac{1}{l+1}}\cdot L_l^1(\eta^2)$. $L_n^m (x)$
is a generalized Laguerre polynomial. In the following this
evolution matrix will be referred to as a $p_{sa}(\theta,\phi)$
pulse.  Note that this does not assume
 the Lamb-Dicke limit, so the extent of the zero-point motion of the harmonic oscillator
 can be a significant fraction of the wavelength of the field.\\
\indent A spin-cyclotron interaction can be set up using a magnetic
field as described in \cite{SM05}
\begin{equation*}
\mathbf{b}(t) = b(x\hat{x}+y\hat{y})\cos(\omega t+\phi)
\end{equation*}
This leads to the Hamiltonian
\begin{equation*}
H'_{sc} = \frac{g\mu_B
b}{2}\sqrt{\frac{\hbar}{2m\tilde{\omega}_c}}(\sigma_+ a_c
e^{-i\phi}+\sigma_- a_c^\dagger e^{i\phi})
\end{equation*}
If $\omega = \omega_{sc} =\omega_s-\omega'_c$ only the levels
$|\!\!\downarrow\! n l\rangle$, $|\!\!\uparrow\! (n-1) l\rangle$ are
connected. In the basis $|\!\!\downarrow\! n l\rangle$,
$|\!\!\uparrow\! (n-1) l\rangle$ the following evolution operator is
thus obtained
\begin{equation}\label{Usc}
U'_{sc} = \begin{pmatrix} \cos(\frac{\sqrt{n}\theta}{2}) &
ie^{i\phi}\sin(\frac{\sqrt{n}\theta}{2}) \\ ie^{-i\phi}
\sin(\frac{\sqrt{n}\theta}{2}) & \cos(\frac{\sqrt{n}\theta}{2})
\end{pmatrix}
\end{equation}
where $\theta = -g\mu_B b\sqrt{\frac{1}{2m\hbar\tilde{\omega}_c}}t$.
In the following this evolution matrix will be referred to as a
$p_{sc}(\theta,\phi)$ pulse.\\
\indent Fig. 1 shows that the three fields connect all the
eigenstates of the spin-axial-cyclotron system.  In addition, the
levels are connected in such a way that the system is eigenstate
controllable \cite{BBR06}. That is, population can be coherently
transferred from any eigenstate to any other eigenstate. For
example, consider the set of eigenstates illustrated in Fig. 1. The
condition for eigenstate controllability is that the pulses of
frequency $\omega_{sa}$, $\omega_s$ and $\omega_{sc}$ must be
applied {\it sequentially}, and not simultaneously. For example, let
us say we want to transfer the $|000\rangle$ state to the
$|111\rangle$ state. We can do so by the sequence:
$p_s(\pi,\phi_1)$, $p_{sa}(\pi,\phi_2)$, $p_{s}(\pi,\phi_3)$,
$p_{sc}(\pi,\phi_4)$, $p_s(\pi,\phi_5)$, where $\phi_1$, $\phi_2$,
$\phi_3$, $\phi_4$ and $\phi_5$ are arbitrary phases. This is
similar in spirit to the eigenstate transfer schemes for trapped
ions \cite{LawEberly}.

\section{Universal quantum computation}
\noindent A universal set of gates consists of the Hadamard, the
$T$-gate and the controlled-NOT gates \cite{NC00}:
\begin{gather*}
T = \begin{pmatrix} 1 & 0 \\ 0 & e^{i\pi/4} \end{pmatrix}, H =
\frac{1}{\sqrt{2}}\begin{pmatrix} 1 & 1\\ 1 & -1 \end{pmatrix} \\
CNOT = \begin{pmatrix} 1 & 0 & 0 & 0\\ 0 & 1 & 0 & 0 \\ 0 & 0 & 0 &
1\\ 0 & 0 & 1 & 0 \end{pmatrix}
\end{gather*}
In the following, we will demonstrate how to implement a universal
set of gates for the
current system by applying the interactions described above.\\
\indent For the spin qubit, the single-qubit gates are easily
implemented. Thus the $T$ gate is performed up to a global phase
factor using the two pulses $p_s(\pi,\pi/8)$, $p_s(\pi,0)$. An
arbitrary phase gate $\left( \begin{smallmatrix} 1 & 0 \\ 0 &
e^{i\varphi}
\end{smallmatrix}\right)$ is in fact performed just by
$p_s(\pi,\varphi/2)$, $p_s(\pi,0)$. The Hadamard gate is obtained
from the pulse sequence $p_s(\pi,-\pi)$, $p_s(\pi/2,\pi/2)$.\\
\indent Implementing a swap gate
\begin{equation*}
SWAP = \begin{pmatrix} 1 & 0 & 0 & 0\\ 0 & 0 & 1 & 0\\ 0 & 1 & 0 &
0\\ 0 & 0 & 0 & 1 \end{pmatrix}
\end{equation*}
between the spin and cyclotron qubits is complicated by the fact
that driving the transitions $|01l\rangle \leftrightarrow
|10l\rangle$ simultaneously drives the transitions $|02l\rangle
\leftrightarrow |11l\rangle$ and population thus leaves the
computational subspace. This problem can, however, be overcome using
composite pulses \cite{CC00}, a method that has already been applied
to for instance trapped ions \cite{GRLBEHSCB03} and trapped
electrons \cite{SM05}. Applying the pulse sequence $p_{sc}(
\pi/\sqrt{2},0)$, $p_{sc}(2\pi/\sqrt{2},\phi_s)$,
$p_{sc}(\pi/\sqrt{2},0)$, where $\phi_s =
\arccos(\cot^2(\pi/\sqrt{2}))$, implements the swapping gate up to a
phase factor of $e^{i\alpha}$ for $|01\rangle$ and $-e^{-i\alpha}$
for $|10\rangle$, where $\alpha \approx -0.8652$. To convert it to a
proper swap gate we suggest the following. First,
$p_s(\pi,\alpha/2)$, $p_s(\pi,0)$ followed by the swapping gate and
subsequently the pulses $p_s(\pi,(\pi-\alpha)/2)$, $p_s(\pi,0)$
implements a swap gate up to a phase of $\pi$ for $|00\rangle$. As
demonstrated in \cite{SM05} the pulse sequence $p_{sc}(\pi,0)$,
$p_{sc}(\pi/\sqrt{2},\pi/2)$, $p_{sc}(\pi,0)$,
$p_{sc}(\pi/\sqrt{2},\pi/2)$ leads to a controlled-phase gate
endowing all states but $|00\rangle$ with a phase of $\pi$. Thus the
phase of
$|00\rangle$ is corrected by applying this controlled-phase gate.\\
\indent Now implementing any single-qubit gate for the cyclotron
qubit is easy, since $I\times S= SW\!AP\cdot (S\times I)\cdot
SW\!AP$, where $S$ denotes an arbitrary single-qubit gate.
Implementing for instance a $T$ gate, however, simplifies since
sandwiching a $T$ gate for the spin qubit between the swapping
sequence and the same sequence with all phases offset
by $\pi$ leads to a $T$ gate for the cyclotron qubit.\\
\indent To implement a controlled-NOT gate between the spin and the
cyclotron qubit, we first implement the controlled-phase gate as
before by the pulse sequence $p_{sc}(\pi,0)$,
$p_{sc}(\pi/\sqrt{2},\pi/2)$, $p_{sc}(\pi,0)$,
$p_{sc}(\pi/\sqrt{2},\pi/2)$. To convert it to a standard $CZ$ gate
\begin{equation*}
CZ = \begin{pmatrix} 1 & 0 & 0 & 0\\ 0 & 1 & 0 & 0\\ 0 & 0 & 1 & 0\\
0 & 0 & 0 & -1 \end{pmatrix}
\end{equation*}
we simply sandwich it between $NOT$ gates on both qubits. Finally,
applying a Hadamard gate to the cyclotron qubit before and after the
$CZ$ gate leads to
a $CNOT$ gate.\\
\indent To implement a swap gate between the spin and the axial
qubits $p_{sa}$ pulses are naturally employed. As before, it is a
problem that driving for instance the transitions $|0n1\rangle
\leftrightarrow |1n0\rangle$ simultaneously drives the transitions
$|0n2\rangle \leftrightarrow |1n1\rangle$ and population thus leaves
the computational subspace. Noting that $\theta_{l=1} =
\frac{1}{\sqrt{2}}(2-\eta^2)\theta_{l=0}$ we can, however, in the
limit where $\eta \approx 0$ (for which $\theta_{l=1}=\sqrt{2}
\theta_{l=0}$) use the composite pulse idea employed for the
spin-cyclotron interaction. If all phases are offset by $-\pi/2$
(cf. formulas \eqref{Usa} and \eqref{Usc}) we can use exactly the
same sequences as for the spin-cyclotron case to implement the
desired gates. Notice that this method can also be applied for a
nonzero $\eta$. For instance, choosing $\eta = 2$ leads to
$\theta_{l=1}=-\sqrt{2}\theta_{l=0}$ and the
pulses above are immediately applicable.\\
\indent A controlled-NOT gate between the cyclotron and the axial
qubits is implemented by the sequence $CNOT_{cyclo,spin}$,
$CNOT_{spin,axial}$, $CNOT_{cyclo,spin}$, $CNOT_{spin,axial}$.\\
\indent So we have now demonstrated how to perform universal
computation with three-qubits in a trapped electron scheme.

\section{Implementing the Deutsch-Jozsa algorithm}
\noindent In this section, we will demonstrate how to implement the
3-qubit Deutsch-Jozsa (DJ) algorithm \cite{DJ92}. The refined version
suggested in \cite{CKH98} allows us to test functions for three
qubits instead of just two with three physical qubits available, as
the control register is eliminated. Testing functions on three
qubits compared to two is interesting, since three is the minimum
number of qubits required to solve the DJ algorithm in a
non-classical way \cite{CKH98}. The algorithm works in the following
way: 1) Initialize the system to $|000\rangle$, 2) Apply a Hadamard
gate to each qubit, 3) Apply the unitary operator $U_f$ defined by
$U_f |x\rangle = (-1)^{f(x)}|x\rangle$, 4) Apply a Hadamard gate to
each qubit, 5) Measure the state of the qubits. If the measurement
returns $|000\rangle$ the function is constant, else it is
balanced.\\
\indent We have already demonstrated how to implement a Hadamard
gate for each qubit, so we only need to implement the $U_f$ gates.
Now, for three qubits there are 2 constant functions and 70 balanced
functions. These functions are indexed as in \cite{KLLC00} using
their outputs expressed as hexadecimal numbers such that for
instance the function with the output $f(0)\cdots f(7) = 00001111$
is denoted $f_{0F}$. Since $U_{f_x} = -U_{f_{FF-x}}$ we only need to
implement 1 constant and 35 balanced functions. The constant
function corresponding to $f_{00}$ is easily implemented since
$U_{f_{00}}$ is equal to the identity. To implement the balanced
functions we refer to \cite{KLLC00}, where NMR sequences for the 35
balanced functions are written down. From this work it follows that
all functions can be constructed from gates on the form
\begin{equation*}
\begin{pmatrix} 1 & 0\\ 0 & -1 \end{pmatrix},
\begin{pmatrix} 1 & 0\\ 0 & \pm i
\end{pmatrix}, \begin{pmatrix} 1 & 0 & 0 & 0\\ 0 & \pm i & 0 & 0\\
0 & 0 & \pm i & 0\\ 0 & 0 & 0 & 1 \end{pmatrix}
\end{equation*}
The single-qubit gates are easily implementable for the spin qubit.
The first one, corresponding to $I_{z}(\pi)$ in \cite{KLLC00}, is
obtained from the sequence $p_s(\pi,\pi/2)$, $p_s(\pi,0)$, whereas
the second one, corresponding to $I_{z}(\pm \pi/2)$, is given by
$p_s(\pi,\pm \pi/4)$, $p_s(\pi,0)$. Using $SWAP$ gates these gates
are transformed to single-qubit gates for the cyclotron and the
axial qubits and thus all necessary single-qubit gates listed in
\cite{KLLC00} are obtained. Likewise the two-qubit operations are
easily achieved. Thus the sequence $CNOT$,$\left(
\begin{smallmatrix} 1 & 0\\ 0 & \pm i
\end{smallmatrix} \right)$, $CNOT$ implements the desired two-qubit
gates, that is, the gates $J_{ij}(\pm \pi/2)$ in \cite{KLLC00}. It
follows that all the $U_f$ functions are implementable, and
therefore that the Deutsch-Jozsa algorithm can be executed in our
system.

\section{Discussion and Conclusion}
\noindent In conclusion, we have demonstrated universal quantum
computation for trapped electron three-qubit states.  We have
explicitly determined a universal set of gates by finding the
necessary pulse sequences. We have also demonstrated that the
three-qubit Deutsch-Jozsa algorithm can be implemented in this system.\\
\indent Thus, this work expands current work in the area of
trapped-electron quantum computing to exploit another degree of
freedom in each electron. Besides constituting an interesting system
in itself in that it can be used to test for example the
Deutsch-Jozsa algorithm, this work could also be useful for studying
many-particle gates via the Coulomb interaction between electrons in
planar Penning traps (the system is not scalable with the trap
considered in this proposal). Each particle could have three qubit
degrees of freedom and one of these could be used for communication.
Such a situation is highly interesting for the
case of error correction. \\
\indent However, there are significant experimental challenges to
the implementation of this proposal at present. The primary
challenge is that the axial motion cannot yet be cooled to the
ground state (which is necessary for initialization). There are
proposals to do so using the cyclotron motion, but these are still
in their early stages. Nevertheless we expect that our work will
motivate rapid experimental progress in this direction, much as
early quantum computing proposals \cite{CZ95} motivated the motional
cooling of trapped {\it ions} \cite{KWMTLIMW98} about a decade ago.
A second challenge is that all the three-qubit states cannot be
measured directly. The spin qubit can be measured
\cite{WABBDHHQSVVV06}, as can the axial qubit with a low temperature
circuit \cite{OHDG}.  For the cyclotron qubit, a scheme adopted in
\cite{CMT01} can be used. In that work, in which the spin and
cyclotron levels are qubits, it is demonstrated that the spin and
cyclotron levels can be coupled with the axial motion and
subsequently the axial motion can be measured to get a cyclotron
qubit measurement. Although this system has long coherence times,
there are notable sources of decoherence that set a limit on the
number of gate operations that can be performed, for instance
fluctuating magnetic and electric fields, thermal noise and
intensity noise from the microwave sources. In spite of these
experimental challenges to be overcome, this system provides an
attractive quantum computing paradigm to be investigated.

\section{Acknowledgements}
\noindent We thank Klaus M{\o}lmer for valuable discussions, and
Chris Monroe for several helpful suggestions.  LHP thanks the
University of Windsor for hospitality.  CR is supported by NSERC,
Canada.

\end{document}